\documentclass[aps,prl,twocolumn,superscriptaddress]{revtex4-2}

\usepackage{graphicx}
\usepackage{amsmath}
\usepackage{mathptmx}
\usepackage{ulem}
\usepackage{xcolor}
\usepackage{orcidlink}
\usepackage[final]{changes}

\begin{document}

\title{\added{A sleight of light: tuning the terahertz response of liquids by creating polar many-body excitations}}

\author{Gerard McCaul}
\email{g.mccaul@lboro.ac.uk}
\affiliation{Department of Physics, Loughborough University, UK}
\thanks{These two authors contributed equally.}

\author{Matthias Runge}
\email{runge@mbi-berlin.de}
\affiliation{Max-Born-Institut f\"ur Nichtlineare Optik und
Kurzzeitspektroskopie, 12489 Berlin, Germany}
\thanks{These two authors contributed equally.}

\author{Michael Woerner}
\affiliation{Max-Born-Institut f\"ur Nichtlineare Optik und
Kurzzeitspektroskopie, 12489 Berlin, Germany}

\author{Diyar Talbayev}
\affiliation{Tulane University, New Orleans, Louisiana 70118, United States}

\author{Thomas Elsaesser}
\affiliation{Max-Born-Institut f\"ur Nichtlineare Optik und
Kurzzeitspektroskopie, 12489 Berlin, Germany}

\author{Denys I. Bondar \orcidlink{0000-0002-3626-4804}}
\affiliation{Tulane University, New Orleans, Louisiana 70118, United States}

\date{\today}

\begin{abstract}
A surprising result from the theory of quantum control is the degree to which the properties of a physical system can be manipulated. Both atomic and many-body solid state models admit the possibility of creating a ‘driven imposter’, in which the optical response of one material mimics that of a dynamically distinct system. Here we apply these techniques to polarons in polar liquids. Such quasiparticles describe solvated electrons interacting with many-body degrees of freedom of their environment. The polaron frequency, which depends on the electron concentration in the liquid, is controlled with a pump field. \added{Using this field as a control, we demonstrate the aforementioned dynamical mimicry experimentally, rendering the polaron frequency of three different liquids identical.}
\end{abstract}

\maketitle

\textit{Introduction:}
Over the previous century, the fundamental principles governing physical systems have (at least in experimentally accessible regimes) been comprehensively uncovered. With a firm grasp of quantum dynamics in hand, a natural question becomes how microscopic behaviour might be \textit{exploited}. In the present computational age, the dominating preoccupation has centred on how physical systems might themselves be forced to compute. The deep links between physical and informational dynamics \cite{kieu_quantum_2003, bondar_uncomputability_2020} means that the \textit{control} of the former represents a route to mastery of the latter. Examples of this premise abound, where quantum control techniques can be employed in the development of variational quantum algorithms \cite{magann_pulses_2021}, combinatorial optimisation \cite{magann_lyapunov-control-inspired_2022}, and single-atom computation  \cite{mccaul_towards_2023}. More broadly, while quantum control has applications in both information processing \cite{glaser_training_2015} and quantum technologies in general \cite{kochControllingOpenQuantum2016}, it also enables investigations of a more fundamental nature. In particular, quantum control is a tool by which the ultimate limits of the malleability and manipulability of matter can be probed. 

 One particularly fruitful approach by which questions of this nature can be answered is \textit{quantum tracking control} \cite{zhu_quantum_2003,gross_inverse_1993, ong_invertibility_1984}. This technique seeks to calculate the driving field required such that the trajectory of some observable expectation conforms to a prespecified ``track'' \cite{rothmanObservablepreservingControlQuantum2005}. In combination with other techniques \cite{sahebiSwitchingOptimalAdaptive2018,mirrahimiReferenceTrajectoryTracking2005,coronQuantumControlDesign2009}, tracking control has successfully been implemented in a variety of contexts for atomic \cite{camposHowMakeDistinct2017}, molecular \cite{magannSingularityfreeQuantumTracking2018,magannQuantumTrackingControl2023,chenCompetitiveTrackingMolecular1995} and solid-state systems \cite{mccaulDrivenImpostersControlling2020,mccaulControllingArbitraryObservables2020,magannSequentialOpticalResponse2022}.

In the case of these latter systems, microscopic models of strongly-interacting systems make explicit the large degree to which the response of driven systems can be tailored \cite{mccaulControllingArbitraryObservables2020}. In particular, they demonstrate a number of intriguing (and potentially exploitable) phenomena. This includes the existence of ``twinning fields'', a driving field which elicits an identical response from two dynamically distinct materials \cite{mccaulOpticalIndistinguishabilityTwinning2021}, and the non-uniqueness of driving fields in generating a given response \cite{mccaulNonUniquenessDriving2022a}. When an observable is a nonlinear function of driving, it is possible to create ``driven imposters'' where a given system's response can be tailored to mimic that of an entirely different material \cite{mccaulDrivenImpostersControlling2020}. This presents an opportunity in both materials science and chemistry to use simpler and cheaper compounds in combination with driving to obtain some property that would otherwise be absent \cite{Zhang2005,10.1021/ja00081a012,10.1021/ar00028a010,C4RA01210K,10.1021/am506611j}.  

It is in this final possibility that tracking control has its most romantic potential application - the realisation of the alchemical dream of making lead look like gold. This objective is however also where the limitations of tracking control models considered to this point reveal themselves. In particular, to derive the driving fields necessary to induce whatever behaviour is desired from a system, a microscopic model linking driving field to observable is required. This together with the fact that these solutions for tracking tend to require broad-band control fields \cite{mccaulControllingArbitraryObservables2020} have hampered experimental validation of the technique. This naturally raises the question of whether these drawbacks might be circumvented by applying the methodology of tracking control to models more tightly bound to observable behaviour in the laboratory. 

This paper answers this question positively, with a novel application of tracking control techniques to an experimental scenario. We focus on manipulating the behavior of polarons -- {many-body quasiparticle excitations} formed through electron-phonon interactions in polar liquids. The models describing {this type of} polaronic response are shown to admit the possibility of ``polar impostorons'' - where distinct liquids may be engineered to exhibit synchronized polaron frequencies {in the terahertz (THz)} range. From this, it is possible to demonstrate experimentally the existence of said polar impostorons. This permits a somewhat scriptural sleight of light: giving water the \textit{appearance} of wine -  if we are permissive enough in our definition of wine to include isopropyl alcohol.

\textit{Creating Polar Impostorons}:
{The polaron quasiparticle was originally introduced to account for the coupling between an electron and excitations of a polar crystal lattice \cite{pekar1946local}, i.e., phonons, and in order to capture many-body aspects of charge transport in polar solids \cite{La33,La48}. For the electron solvated in a polar liquid - a prototypical quantum system \cite{marsalekStructureDynamicsReactivity2012,turiTheoreticalStudiesSpectroscopy2012,herbertStructureAqueousElectron2019} -the polaronic theory of an electron in a polarizable medium has been employed to describe its quasi-static localization potential, quantum states, and single-particle excitations \cite{Jo59,We60,We72,Ma87,La91}.} Descriptions of this type have been deployed in a wide number of contexts \cite{crevecoeur1970electrical, mckenna2012two,de2016tracking, Bruderer07, coropceanu2007charge}, as polarons play a key role in a broad range of phenomena \cite{Franchini2021}.

{Of particular relevance to the present work is the presence of polarons in dynamics of hydrated electrons \cite{laria_reference_1991, PhysRevB.46.9958}.Recent experimental work has established the dynamic many-body response of solvated electrons, which originates from the Coulomb coupling of electronic and longitudinal nuclear degrees of freedom of the liquid environment and, thus, bears a polaronic character as well \cite{GH21B,SI22B,WO22,Runge2023}.} After (nonlinear) multiphoton ionization of solvent molecules by a femtosecond optical pulse, the subpicosecond relaxation of the generated free electrons into their solvated ground state launches long-lasting coherent polaron oscillations impulsively and with a frequency $\nu_0$ determined by the electron concentration $c_e$. Similar to the impact of polarons in solids \cite{gaalInternalMotionsQuasiparticle2007,leeMotionSlowElectrons1953,frohlichElectronsLatticeFields1954}, the dielectric response of the liquid is modified and a polaron resonance arises at $\nu_0$ in the terahertz (THz) frequency range.
 

Despite the fact that this scenario implies microscopic many-body, strongly interacting dynamics, the frequency of the polaronic response is defined by the zero point of the real part of the liquid's dielectric function Re$(\epsilon (\nu_0, c_e))= \epsilon' (\nu_0, c_e)=0$ \cite{GH21B}. On a macrosopic level, this is well described by the local-field model of Clausius and Mossotti, in which electrons and solvent molecules are approximated as interacting point-like dipoles, leading to \cite{HA83A,KU57A}:
\begin{eqnarray}
	3\dfrac{\epsilon(\nu,c_e)-1}{\epsilon(\nu,c_e)+2}&=&3\dfrac{\epsilon_\text{neat}(\nu)-1}{\epsilon_\text{neat}(\nu)+2} + c_eN_A\alpha_\text{el}(\nu),
	\label{eq:ClausiusMossotti}\\
	\text{with  }\alpha_\text{el}(\nu)&=&-\dfrac{e^2}{\epsilon_0m[(2\pi\nu)^2+i\gamma(2\pi\nu)]} 
	\label{eq:ElectronPolarization} \\
	&{\overset{\gamma=0}{=}}&-\dfrac{e^2}{\epsilon_0m(2\pi\nu)^2}
	\label{eq:alphareal},
\end{eqnarray}

Here $\epsilon(\nu,c_e) = \epsilon'(\nu,c_e)+i \epsilon''(\nu,c_e)$
and $\epsilon_\text{neat}(\nu)=\epsilon_\text{neat}'(\nu) + i \epsilon_\text{neat}''(\nu)$
 is the dielectric function of the liquid containing solvated electrons and of the neat liquid, respectively; $N_A$ is Avogadro's constant and  $\alpha_\text{el}$ is the polarizability of a solvated electron with the elementary charge  $e$ and the electronic mass $m$. The damping time of the polaron oscillations in alcohols is on the order of $T_2 \geq 100$ ps and, thus, the local friction rate $\gamma$ is small compared to the damping of $\epsilon_\text{neat}$. In the following, we set $\gamma = 0$, making the polarizability $\alpha_\text{el}$ a real quantity (Eq.~\eqref{eq:alphareal}). The line shape of the polaron resonance is given by
$-\text{Im}[\epsilon(\nu)^{-1}]=\epsilon'(\nu)[\epsilon'(\nu)^2+\epsilon''(\nu)^2]^{-1}$.

The polaron frequency $\nu_0$ depends on $\epsilon(\nu,c_e)$ and has been tuned over a frequency range of some 5 octaves (0.2 to 6 THz) by changing the electron concentration $c_e$ in water and different alcohols \cite{GH21B,SI22B,Runge2023}. This indirect mechanism for manipulating polaronic behaviour via  $c_e$ is itself directly controllable by the femtosecond ionization pulse. In this sense, it is possible to take two liquids with distinct dielectric functions, and drive them such that their polaron frequency becomes identical. Within this spectrally limited context, they become indistinguishable. In the spirit of \cite{mccaulDrivenImpostersControlling2020}, we refer to these frequency aligned quasiparticles as \textit{polar impostorons}.

To analyze this scenario, we first invert Eq.~\eqref{eq:ClausiusMossotti} to express the electron concentration in terms of the dielectric functions and polarizability:
\begin{equation}
	c_e=\dfrac{3}{N_A\alpha_\text{el}(\nu)}\left(\dfrac{\epsilon(\nu,c_e)-1}{\epsilon(\nu,c_e)+2}-\dfrac{\epsilon_\text{neat}(\nu)-1}{\epsilon_\text{neat}\nu)+2}\right). \label{eq:ce}
\end{equation} 
with $\alpha_\text{el}$ given by Eq.~\eqref{eq:alphareal}. At the frequency $\nu = \nu_0$, $\epsilon'(\nu_0,c_e)=0$, leading to
\begin{equation}
	c_e=\dfrac{3}{N_A\alpha_\text{el}(\nu_0)}\left(\dfrac{i\epsilon''(\nu_0,c_e)-1}{i\epsilon''(\nu_0,c_e)+2}-\dfrac{\epsilon_\text{neat}(\nu_0)-1}{\epsilon_\text{neat}(\nu_0)+2}\right). \label{eq:ce_v0}
\end{equation}  
or, after a separation of the real and imaginary parts,
\begin{eqnarray}
	\text{Re}(c_e)&=&\dfrac{3}{N_A\alpha_\text{el}(\nu_0)}\left(\dfrac{\epsilon''(\nu_0,c_e)^2-2}{\epsilon''(\nu_0,c_e)^2+4}-\right.\nonumber\\ &&\hspace{2.5cm}\left. \dfrac{\Sigma_\text{neat}+\epsilon'_\text{neat}(\nu_0)-2}{\Sigma_\text{neat}+4\epsilon'_\text{neat}(\nu_0)+4}\right) \label{eq:re_ce}\\
	\text{Im}(c_e)&=&\dfrac{3}{N_A\alpha_\text{el}(\nu_0)}\left(\dfrac{3\epsilon''(\nu_0,c_e)}{\epsilon''(\nu_0,c_e)^2+4}-\right.\nonumber\\ &&\hspace{2.5cm}\left.\dfrac{3\epsilon''_\text{neat}(\nu_0)}{\Sigma_\text{neat}+4\epsilon'_\text{neat}(\nu_0)+4}\right)\label{eq:Imce}
\end{eqnarray}
with $\Sigma_\text{neat}=|\epsilon_\text{neat}(\nu_0)|^2$. Since the electron concentration $c_e$ is a real quantity (i.e. Im$(c_e)=0$), equation \eqref{eq:Imce} leads to 
\begin{equation}
	\dfrac{\epsilon''(\nu_0)}{\epsilon''(\nu_0)^2+4}=\dfrac{\epsilon''_\text{neat}(\nu_0)}{\Sigma_\text{neat}+4\epsilon'_\text{neat}(\nu_0)+4}.
	\label{eq:Ime_vs_Imneat}
\end{equation}
showing that the imaginary part $\epsilon''(\nu_0)$ is entirely determined by the dielectric function of the neat liquid and, consequently, independent from the electron concentration $c_e$. 

We now return to creating impostorons by making the polaron frequencies $\nu_0$ of two different liquids $(1)$ and $(2)$ identical by adjusting the electron concentrations $c_e^{(1)}$ and $c_e^{(2)}$ in a way that $\epsilon'_1(\nu_0) = \epsilon'_2(\nu_0)=0$. Here the index $(i=1,2)$ in $\epsilon_i$ designates the respective liquid. Eq.~\eqref{eq:re_ce} gives the following relation for the concentration difference $c_e^{(1)} - c_e^{(2)}$:
\begin{eqnarray}
	c_e^{(1)}-c_e^{(2)}&=&\dfrac{3}{N_A\alpha_\text{el}(\nu_0)} \left[\dfrac{\epsilon''_1(\nu_0)^2-2}{\epsilon''_1(\nu_0)^2+4} - \dfrac{\Sigma_\text{neat}^{(1)}+\epsilon'_\text{1,neat}(\nu_0)-2}{\Sigma_\text{neat}^{(1)}+4\epsilon'_\text{1,neat}(\nu_0)+4}\right. \label{eq:control} \nonumber \\
	&&\left.- \dfrac{\epsilon''_2(\nu_0)^2-2}{\epsilon''_2(\nu_0)^2+4}+\dfrac{\Sigma_\text{neat}^{(2)}+\epsilon'_\text{2,neat}(\nu_0)-2}{\Sigma_\text{neat}^{(2)}+4\epsilon'_\text{2,neat}(\nu_0)+4}\right] 
\end{eqnarray}
Eq.~\eqref{eq:control} shows that an appropriate choice of the two electron concentrations results in an identical polaron frequency $\nu_0$.  The exact value of $\nu_0$ can be tuned via the absolute values of the two concentrations, while Eq.~\eqref{eq:control} sets the concentration {\it difference}. That is, setting the difference guarantees an identical response, while the individual concentrations determine the frequency at which this occurs. Notably, Eq.~\eqref{eq:Ime_vs_Imneat} is fulfilled at $\nu_0$ only, and identical polaron frequencies are not sufficient to imply an identical frequency dispersion of the dielectric functions of the two liquids around $\nu_0$. 
 
In general, the two imaginary parts $\epsilon''_{1,2}(\nu)$, which govern the polaronic line shape $-\text{Im}[\epsilon_i(\nu)^{-1}]$ are different, though their absolute value in the THz frequency range is small for liquids such as alcohols. In the vicinity of $\nu_0$, the  dielectric function $\epsilon_i(\nu,c_e^{(i)})$ of liquid (i) can be well approximated by


\begin{eqnarray}
\epsilon_i(\nu, c_e^{(i)})&\approx & B_i(c_e^{(i)}) (\nu - \nu_0) + i \epsilon_i''(\nu_0)\nonumber\\
B_i(c_e^{(i)}) &=& \left[\frac{\partial \epsilon'_i(\nu, c_e^{(i)})}{\partial \nu}\right]_{\nu = \nu_0} \\
-\text{Im}[\epsilon_i(\nu, c_e^{(i)})^{-1}]) &\approx& \frac{1}{\epsilon''_i(\nu_0)} \left[1+ \frac{[B_i(c_e^{(i)}) (\nu - \nu_0)]^2}{\epsilon''_i(\nu_0)^2} \right]^{-1} 
\end{eqnarray}
Thus, the spectral profile of the polaron resonance $-\text{Im}[\epsilon_i(\nu, c_e^{(i)})^{-1}]$ in two different liquids can be made similar by finding an appropriate $\nu_0$ for which the relation
\begin{eqnarray}
\frac{B_1(c_e^{(1)})}{\epsilon''_1(\nu_0)}&=& \frac{B_2(c_e^{(2)})}{\epsilon''_2(\nu_0)}\label{eq:profile}
\end{eqnarray} 
is fulfilled in addition to the condition of Eq.~\eqref{eq:control} mentioned above. This means that if one wishes to match only the polaron frequency (given by $\epsilon'(\nu_0, c_e)=0$), then only the difference on concentrations is constrained, and  $\nu_0$ can be tuned over a range. If however we also wish to match the spectral profile, then Eq.~\eqref{eq:profile} typically constrains the systems such that only a single $\nu_0$ satisfies both relations.

\textit{Experimental Results}: For an experimental demonstration of the impostoron concept, we studied the polaronic response {of a prototypical selection of liquids, namely 2-propanol (isopropanol, IPA), ethylene glycol (EG), and water. Such liquids display a different molecular arrangement and hydrogen bond structure. Their polarity and, thus, static dielectric constants cover a wider range with values of $\epsilon_\text{static}=17.9$ for IPA, 37 for EG, and 81 for water. }
 In the experiments,  a gravity driven liquid jet with a thickness of about 100~$\mu$m served as a sample. 
 
 Electrons were generated by multiphoton ionization of solvent molecules with optical pulses of a center wavelength of 800~nm, a pulse energy of $65-120~\mu$J, a pulse duration of 66~fs and a repetition rate of 1~kHz. The excitation pulses are focused with a lens (focal length 1~m) to a spot size of about 200~$\mu$m on the liquid. At a variable delay time $\tau$ after electron generation, the  liquid jet is probed with a THz pulse generated by difference-frequency mixing in a MgO-doped LiNbO$_3$ crystal. The THz pulse has a center frequency of 0.7~THz, a spectral width of 0.5 THz (FWHM) and is focused to a spot size of about 700~$\mu$m on the sample by an off-axis parabolic mirror (focal length 25.4~mm). The electric field of the THz pulse transmitted through the sample is detected as a function of real time $t$ by free-space electrooptic sampling. A mechanical chopper with a frequency of 500~Hz in the beam path of the excitation pulses allows for subsequent measurements of the THz probe electric fields  $E^\text{pumped}_\text{Pr}(t,\tau)$ and $E_\text{Pr}(t,\tau)$ with and without solvated electrons present. The nonlinear THz response of the photoexcited polar liquid is derived as $E_\text{NL}(t,\tau)=E^\text{pumped}_\text{Pr}(t,\tau)-E_\text{Pr}(t,\tau)$. Further details of the experimental method have been given in Refs.~\cite{GH21B,SI22B}.

\begin{figure}[t!]
	\begin{center}
		\includegraphics[width=0.5\textwidth]{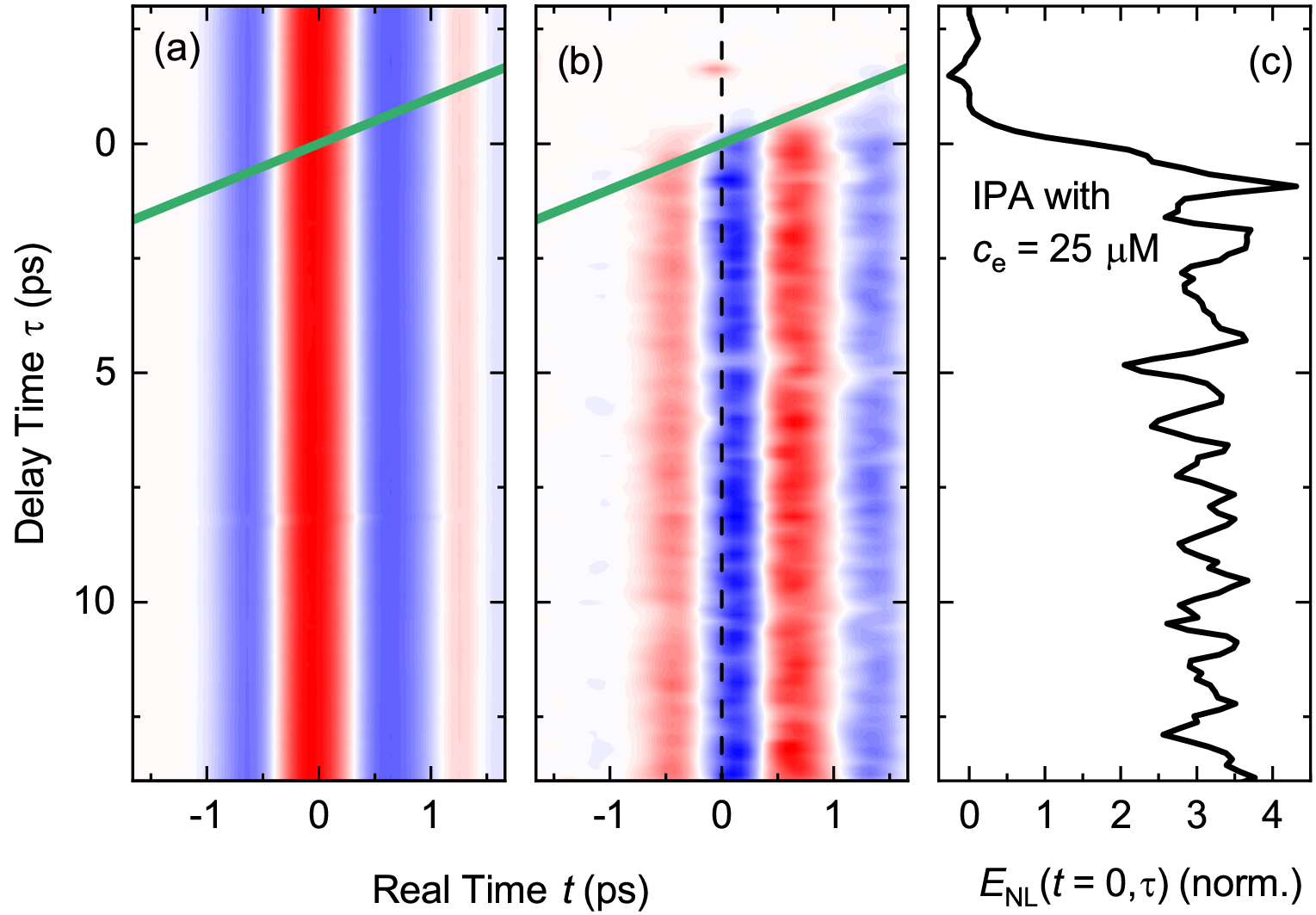}
	\end{center}
	\caption{(a)~THz probe electric field $E_\text{Pr}(t,\tau)$ as a function of real time $t$ (abscissa) and delay time $\tau$ (ordinate) and temporal position of an intense 800-nm  pulse (green line) to generate electrons via multiphoton ionization. (b)~Nonlinear THz response $E_\text{NL}(t,\tau)$ as a function of real time $t$ and delay time $\tau$ for IPA with an electron concentration of $c_e=25~\mu$M.  (c)~Cut along delay time $\tau$ at the maximum of the nonlinear THz response $E_\text{NL}(t,\tau)$ indicated by the dashed line in panel~(b).}   
	\label{fig:exp_IPA}
\end{figure}

 Fig.~\ref{fig:exp_IPA} presents experimental results for an electron concentration of $c_e=25~\mu$M in IPA. In Fig.~\ref{fig:exp_IPA}(a), the THz probe pulse $E_\text{Pr}(t,\tau)$ transmitted through the unexcited liquid jet is plotted as a function of real time $t$ and delay time $\tau$. The green line indicates the temporal position of the femtosecond pump pulse generating electrons.
Figure~\ref{fig:exp_IPA}(b) shows the nonlinear THz response $E_\text{NL}(t,\tau)$ of the excited sample. The data were filtered applying a 2D Fourier filter with a bandwidth of 4~THz in the 2D spectral domain. $E_\text{NL}(t,\tau)$ rises after excitation at $\tau=0$ with a step-like character,  due to a sudden absorption change within the 0.5 THz bandwidth of the THz probe pulse after  electron generation. Figure~\ref{fig:exp_IPA}(c) displays  a cut along delay time $\tau$ through the maximum of $E_\text{NL}(t,\tau)$ at $t=0$, which exhibits pronounced oscillations superimposed with the step-like signal. These oscillations  arise from coherent polaron oscillations, impulsively excited during the ultrafast relaxation of the photogenerated electrons to a localized ground state.

The polaron consists of an electron which couples via the Coulomb interaction to electronic and nuclear degrees of freedom of the surrounding liquid. The relevant volume of the liquid is roughly set by the (Debye) screening length of Coulomb interaction and can be represented by a sphere of a nanometer diameter \cite{SI22B}, containing up to tens of thousands of solvent molecules. The  polaron oscillations represent longitudinal modulations of space charge within this volume leading to oscillations of the polaron size,  which modify the transverse dielectric function. This mechanism makes the polaron oscillations accessible to optical probing with transverse electric fields. The polaron frequency $\nu_0$ is set by $\epsilon' (\nu_0, c_e)=0$ of the  longitudinal macroscopic dielectric function \cite{FU68A}, which in the THz frequency range is almost identical to the transverse macroscopic dielectric function \cite{BO98A}. 

\begin{figure}[t!]
	\begin{center}
		\includegraphics[width=0.5\textwidth]{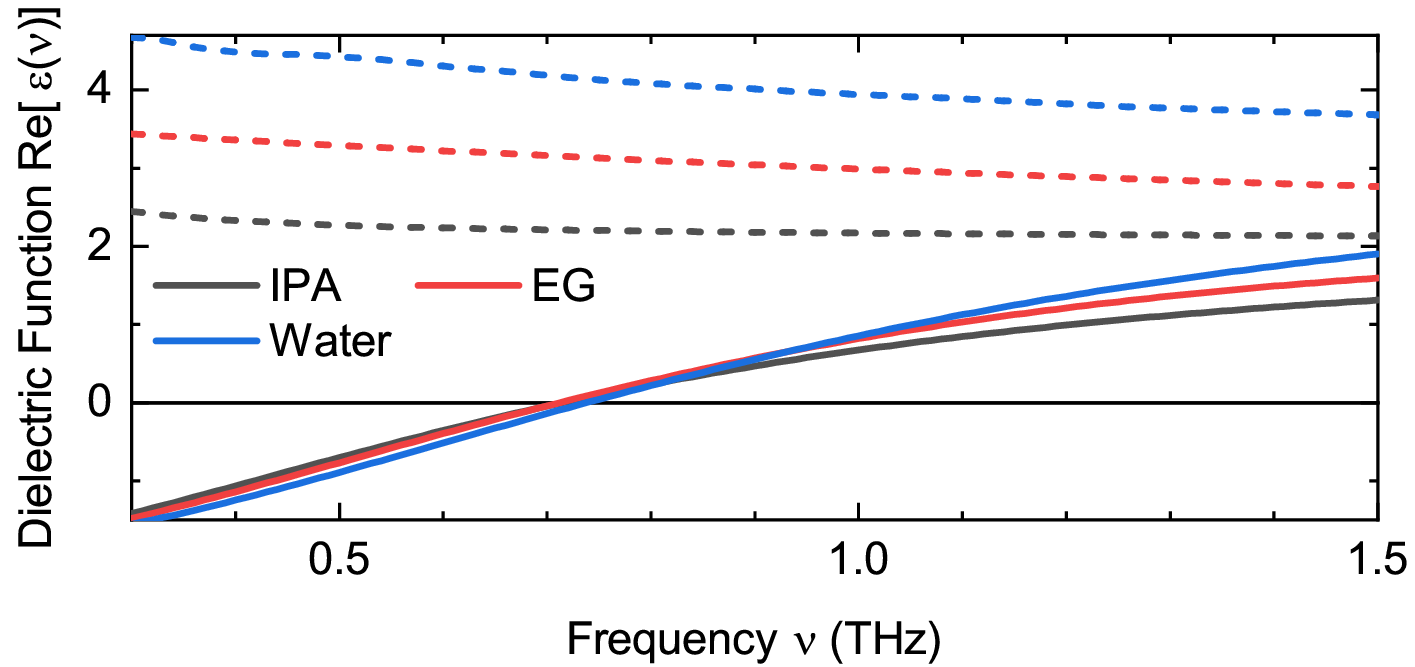}
	\end{center}
	\caption{Real part of the dielectric function of IPA, EG and water without electrons (dashed lines) and with an electron concentration of $c_e=25$, 30 and 40~$\mu$M calculated from the Clausius Mossotti equation (solid lines). }   
	\label{fig:CM_real_epsilon}
\end{figure} 
In different polar liquids, the real parts Re($\epsilon_i(\nu,c_e))= \epsilon'_i(\nu,c_e)$ of the macroscopic dielectric functions according to Eq.~\eqref{eq:ClausiusMossotti} become very similar for appropriate choices of electron concentrations $c_e$, as illustrated in Fig.~\ref{fig:CM_real_epsilon} for electron concentrations of $c_e=25$, 30 and 40$~\mu$M in IPA, EG and water. All three curves display a zero crossing at about $\nu_0 = 0.7$~THz and very similar dispersion in the low-frequency THz range.

\begin{figure}[t!]
	\begin{center}
		\includegraphics[width=0.5\textwidth]{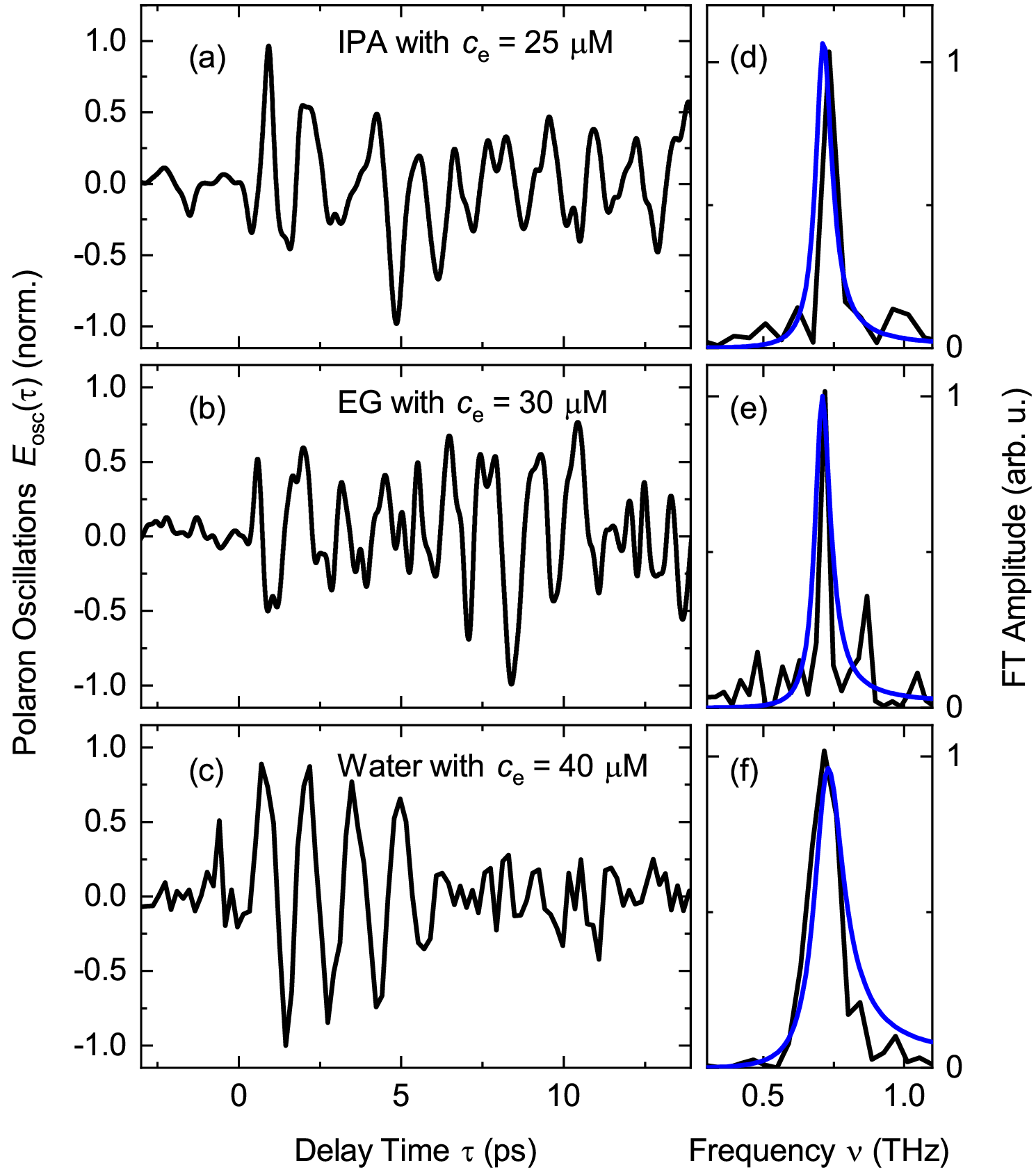}
	\end{center}
	\caption{Making the ultrafast polaronic THz response of solvated electrons in the polar liquids IPA, EG and Water indistinguishable. IPA, EG and water have electron concentrations of $c_e=25$, 30, and 40$~\mu$M. (a-c)~Pure oscillatory signals derived by a cut through the nonlinear signal $E_\text{NL}(t,\tau)$ at $t=0$ along delay time $\tau$.  (d-f)~Fourier transforms of the oscillatory signals in panels~(a-c) with maxima at an oscillation frequency of about 0.7~THz. The blue curves give the spectra $-\text{Im}[\epsilon(\nu)^{-1}]$ with the macroscopic dielectric function $\epsilon(\nu)$ calculated by Eq.~\eqref{eq:ClausiusMossotti}. }   
	\label{fig:exp_comp}
\end{figure}

Figure~\ref{fig:exp_comp}(a) shows the polaron oscillations in  IPA extracted by subtracting the step-like contribution from the cut in Fig.~\ref{fig:exp_IPA}(c). In Figs.~\ref{fig:exp_IPA}(b,~c), the same procedure was applied to equivalent data sets measured in EG and water. All three transients are normalized to their maximum and show clear oscillations for delay times $\tau>0$, which span in the cases of IPA and EG [cf. Figs.~\ref{fig:exp_comp}(a,~b)] the whole time range, reflecting the highly underdamped character of the polaron oscillations. {The Fourier transforms of the oscillations  [Figs.~\ref{fig:exp_comp}(d-f)]} display their maxima at an identical polaron frequency $\nu_0 = 0.7$~THz. The linewidths and center frequencies are well reproduced by the imaginary part of the reciprocal dielectric function $-\text{Im}[\epsilon_i(\nu)^{-1}]=\epsilon''_i(\nu)[\epsilon'_i(\nu)^2+\epsilon''_i(\nu)^2]^{-1}$ of the neat liquids around $\nu_0$ (cf. Eq. \ref{eq:Ime_vs_Imneat}). The widths of the spectra are {much larger than a potential linewidth caused by the damping $\gamma$ of the electron polarizability $\alpha_e$ (Eq.\ref{eq:ElectronPolarization}).}

{ For the current choice of electron concentrations $c_e$ in IPA and EG, very similar line shapes of the polaron resonances [blue curves in Figs.~\ref{fig:exp_comp}(d) and (e)] are observed,  i.e., relation \eqref{eq:profile} is approximately fulfilled. The substantially larger linewidth found for water originates mainly from the faster damping  of polaron oscillations on a time scale of about 5~ps. In contrast to IPA and EG, intermolecular hydrogen bond modes and librations contribute to the dielectric spectrum of water in this spectral range and open an additional relaxation channel for the polaron oscillations. As a result, relation \eqref{eq:profile} cannot be fulfilled for water and one of the alcohols.  }

\textit{Conclusions and Outlook}: The results presented here demonstrate the degree to which polaron response can be controlled both by a femtosecond nonlinear generation of free electron density (setting frequency),  and THz probe pulse (controlling amplitude). {Thus, it is possible to create polar impostorons by rendering the dynamic polaron response frequency of distinct liquids  indistinguishable. Beyond this controllability is damping, which manifests in the imaginary part of the dielectric function and originates from the coupling of the collective polaron excitations to other low-frequency modes in the very same frequency range. On the other hand, the small imaginary part of the THz dielectric function in alcohols can be made nearly identical in the frequency range between 1.0 and 1.5 THz, as is evident from the linear THz spectra reported in Ref.~\cite{MA24}. The focus on this regime is due to the fact that it corresponds to the spectral range in which the `molecular fingerprint' of many chemical and biological materials is located, and as such the THz range responses lends itself to a number of non-destructive sensing and imaging applications \cite{el2013review, pickwell2006biomedical, yu2019medical, bessou2012advantage, cao2021characterization,liu2016discrimination, xu2015discrimination, SO1, SO2}.}

{
Furthermore, there is no reason (subject to the same conditions regarding damping), that the impostoron concept should not be applicable to electrons solvated in a wider range of polar liquids, including, e.g., ammonia, amines and others. The caveat to this point is that while it is always possible to manipulate the zero of the real part of the solvent dielectric function (and hence the polaron frequency) through $c_e$, this does not offer independent control the imaginary part, and therefore the line shape of the polaron resonance is not directly controllable. An important future avenue of investigation is therefore whether there exists a secondary control that can influence a parameter such as (for example) the electron damping $\gamma$, and hence offer independent control of both the real and imaginary parts of the dielectric function.} 

Until this point ``driven imposters''~\cite{mccaulDrivenImpostersControlling2020} have been considered a theoretical curiosity, however this work marks a potential new direction in experimental quantum technology by demonstrating the unique possibilities predicted by quantum tracking control.
The present results on polaron dynamics suggest that they are good candidates for the realisation of other exotic optical effects.   In particular, the tracking control formalism has recently been deployed to \textit{dynamically} generate an epsilon near-zero (ENZ) response \cite{WuENZPhotonics, ENZPhotonics} in a material \cite{2301.12069}. An immediate future extension of the work presented here is therefore to apply it to this scenario, dynamically generating ENZ-like responses in polaronic systems. {Finally, the tunability of physical properties as demonstrated here is a prerequisite for a system to serve as a novel platform for computing. Tracking control previously motivated the development of a single-atom computing protocol \cite{mccaul_towards_2023}, and in this context, it is natural to suggest that polaronic control might be used analogously to realise a (literal) form of liquid computing.}


\paragraph{Acknowledgments:}
T.E. and D.I.B. are grateful to the Alexander von Humboldt Foundation for organizing the  11th Humboldt Award Winners’ Forum, where this collaboration was conceived. D.I.B. was supported by Army Research Office (ARO) (grant W911NF-23-1-0288; program manager Dr.~James Joseph), while G.M. is supported by the European Research Council (ERC) under the European Union’s Horizon 2020 Research and Innovation Program (grant agreement 833365). The views and conclusions contained in this document are those of the authors and should not be interpreted as representing the official policies, either expressed or implied, of ARO or the U.S. Government. The U.S. Government is authorized to reproduce and distribute reprints for Government purposes notwithstanding any copyright notation herein. We thank A. Ghalgaoui, J. Zhang, and P. Singh for their contributions to the experimental work. This research has received funding from the European Research Council (ERC) under the European Union’s
Horizon 2020 Research and Innovation Program (Grant No. 833365). D.I.B. and D.T. are also supported by by the W. M. Keck Foundation.

G.M. and M.R. contributed equally to this work.

 \ifx\unpublished\@undefined\def\unpublished{unpublished}\fi

\end{document}